\documentclass[a4paper]{jpconf}

\usepackage{graphicx}
\usepackage{booktabs}
\usepackage{multirow}
\usepackage{units}
\usepackage{doi}
\usepackage{hyperref}

\newcommand{\prog}[1]{\textsc{#1}}
\newcommand{\comm}[1]{\texttt{#1}}
\newcommand{\comp}[1]{#1}
\newcommand{\serv}[1]{#1}
\newcommand{\lang}[1]{#1}
\newcommand{\oper}[1]{#1}

\newcommand{\Cppeleven}{\mbox{C\texttt{++}11}}

\graphicspath{{./images/}}

\usepackage{relsize}
\def\babar{\mbox{\slshape B\kern-0.1em{\smaller A}\kern-0.1em
		B\kern-0.1em{\smaller A\kern-0.2em R}}}

\begin{document}
\title{GooFit 2.0}

\author{Henry Schreiner\textsuperscript{1},
	Christoph Hasse\textsuperscript{2},
	Bradley Hittle\textsuperscript{3},
	Himadri Pandey\textsuperscript{1},
	Michael Sokoloff\textsuperscript{1}\ and
	Karen Tomko\textsuperscript{3}
}

\address{\textsuperscript{1} University of Cincinnati, 2600 Clifton Ave, Cincinnati, OH 45220, USA}
\address{\textsuperscript{2} Technical University of Dortmund, Emil-Figge-Stra\ss e 50, 44227 Dortmund, Germany}
\address{\textsuperscript{3} Ohio Supercomputer Center, 1224 Kinnear Rd, Columbus, OH 43212, USA}

\ead{henry.schreiner@uc.edu}

\begin{abstract}
The \prog{GooFit} package provides physicists a simple, familiar syntax for manipulating probability density functions and performing fits, and is highly optimized for data analysis on \comp{NVIDIA} GPUs and multithreaded CPU backends. \prog{GooFit} was updated to version 2.0, bringing a host of new features. A completely revamped and redesigned build system makes \prog{GooFit} easier to install, develop with, and run on virtually any system. Unit testing, continuous integration, and advanced
    logging options are improving the stability and reliability of the system. Developing new PDFs now uses standard \lang{CUDA} terminology and provides a lower barrier for new users. The system now has built-in support for multiple graphics cards or nodes using \prog{MPI}, and is being tested on a wide range of different systems. 
GooFit also has significant improvements in performance on some GPU architectures due to optimized memory access. Support for time-dependent four-body amplitude analyses has also been added.
\end{abstract}

\section{Introduction}


Multidimensional fits to large datasets, with tens of millions of events, are becoming increasingly common in High Energy Physics (HEP). The models can be computationally intense, with hundreds of parameters. In preparing a model and fitting the data, the most natural description is usually to prepare components,%
\footnote{Although commonly called PDFs, these components include complex quantum mechanical amplitudes whose magnitude-squared values represent PDFs.}
and then combine these components for a specific fit. This is the approach taken by \prog{RooFit} \cite{lib:RooFit}, one of the most commonly used physics fitting packages. With the advent of highly parallel computing architectures, the need arose for a system that would make implementing fitting efficiently in parallel as intuitive  to physicists as 
\prog{RooFit}.


Graphical Processing Units (GPUs)  provide an alternative to traditional computing, with the ability to do computations in a massively parallel compute engine, with thousands of floating point operations processed at the same time (see Table~\ref{tab:specs}). Since most of the time spent in fitting a probability distribution to a dataset is composed of independently computing the value of a function at each point in the dataset, this maps well to the processing capabilities of GPUs.

There are several complications however; programming on a graphics card uses a different language (such as \lang{CUDA} for \comp{NVIDIA}) and requires data to be managed between memory in the host and the device. GPUs have a simpler instruction set, no branch prediction, and an emphasis on single precision compute, especially for the cheaper ``gamer''-class cards. They also must compute the exact same instruction across groups of data at the same time. Even a simple if statement will force a mask to be applied and both sides of the if statement must be computed separately.
These factors combine to make writing GPU code substantially different from traditional CPU code.

\begin{table}
    \caption{\label{tab:specs}The advertised floating point operations per second (FLOP/s) for several common graphics cards, for both single precision (SP) and double precision (DP), and the number of Streaming Multi-processors (SMPs). Some approximate FLOPs of \comp{Intel} CPUs for comparison include \unit[0.0864]{GFLOP/s} for a 2015 i5 dual core processor or  \unit[0.461]{GFLOP/s} for a 12 core Xeon from a similar year.}
\vspace{.1in}
\lineup
\begin{center}
	\begin{tabular}{rllllll}
		\toprule
	&               &                            &                  &           \multicolumn{2}{c}{\textbf{GFLOP/s}} &            \\
	& \textbf{Name} & \textbf{Cores} &  \textbf{Clock}  & \textbf{SP}   & \textbf{DP}   &  \textbf{Cost} \\
	\cmidrule{2-7}
	\multirow{2}{*}{\textbf{Gamer}}  &  GTX 1050 Ti  &           \0768             & \unit[1290]{Mhz} &      \01980       &   \0\0\062    &     \0\0\$150     \\
	&  GTX 1080 Ti  &           3584            & \unit[1596]{Mhz} &      11300       &     \0\0330     &     \0\0\$850     \\
	\cmidrule{2-7}
	\multirow{2}{*}{\textbf{Server}} &   Tesla K40   &                               2880       & \unit[\0745]{Mhz} &   \04290 &  \01430     &    \0\$3000    \\
	&  Tesla P100   &           3584            & \unit[1329]{Mhz} &             \09300  & \04700 &  \$10000\\
	\bottomrule
\end{tabular}
\end{center}
\vspace{-.2in}
\end{table}


One of the first fitting systems designed for parallel fits on a GPU was the \prog{GooFit} package, first released in 2013 by Rolf Andreassen \cite {lib:GooFit:main}. The project was designed using the \lang{CUDA} language from \comp{NVIDIA}, and used their \prog{Thrust} library \cite{lib:Thrust} to manage kernel launches. By the end of the year, \prog{GooFit} was expanded with an \prog{Thrust} supported \lang{OpenMP} backend to enable parallel fitting on CPU devices ``with the flip of a (compiler) switch''. 

\begin{figure}
	\begin{minipage}[b]{2in}
		\begin{center}
			\includegraphics[width=\textwidth]{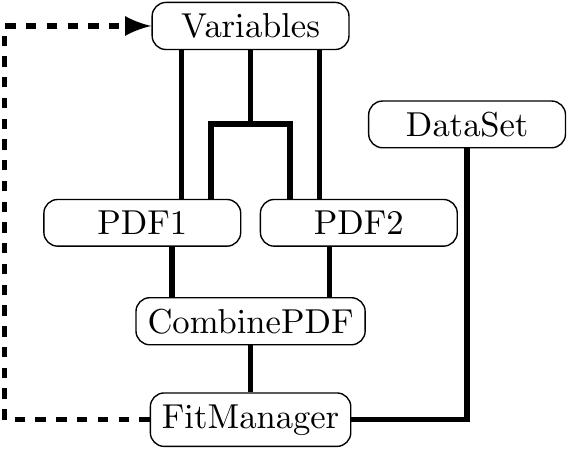}
		\end{center}
	\end{minipage}
	\hspace{.2in}
	\begin{minipage}[b]{4in}
        \caption{\label{fig:fitfeatures}Illustration of the structure of a \prog{GooFit} program. The \comm{Variable}s represent both observables and parameters, and are input to PDFs. The PDFs can be combined using combination PDFs. The final PDF is fed into a \comm{FitManager}, which also takes a binned or unbinned \comm{DataSet} with data to attach to an observable. The \comm{FitManager} performs a fit, and sets the parameter variables accordingly.}
	\end{minipage}
\vspace{-.3in}
\end{figure}

\begin{figure}
	\begin{minipage}[b]{3in}
		\begin{center}
			\includegraphics[width=3in]{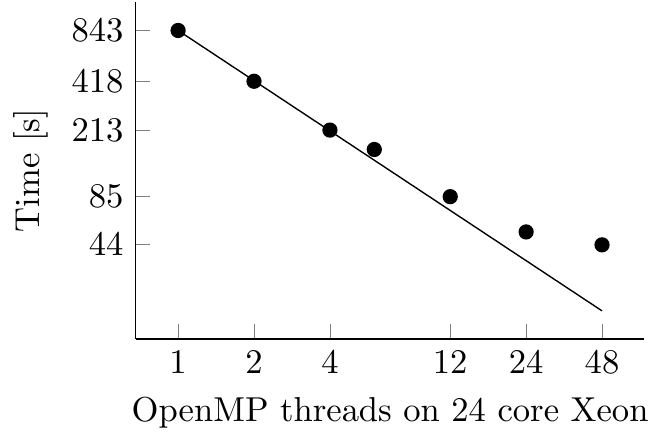}
			\vspace{-.3in}
		\end{center}
        \caption{\label{fig:pipipi}Performance of $\pi^{+}\pi^{-}\pi^{0}$ with 16 time dependent amplitudes on an \comp{Intel} Xeon E5-2680 dual-chip system (24 cores). The line illustrates ideal scaling.}
	\end{minipage}
	\hspace{.2in}
	\begin{minipage}[b]{3in}
		\begin{center}
			\includegraphics[width=3in]{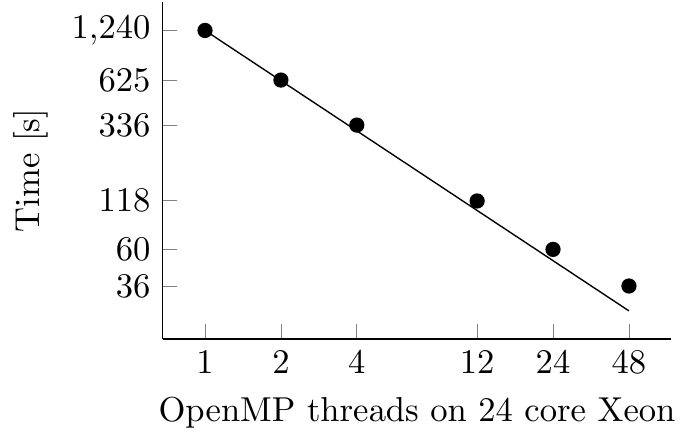}
			\vspace{-.3in}
		\end{center}
		\caption{\label{fig:dmix}Performance of the Zach fit ($D^{*+}$ and $D^{0}$ mass difference measurement with BaBar data) with 16 time dependent amplitudes.}
	\end{minipage}
\vspace{-.2in}
\end{figure}

\prog{GooFit} 1.0 had an impressive list of built-in PDFs, including a specialized system for amplitude analyses of three body particle decays (often called Dalitz plot analyses). Combination PDFs provided ways to build more complex PDFs out of simpler building blocks. Several examples were provided. Simple composition of PDFs using existing building blocks was \prog{GooFit}'s design target (see Figure~\ref{fig:fitfeatures}), and it succeeded in that. The comparative performance for different systems can be seen in Table~\ref{tab:perf}, and scalability when changing the number of \lang{OpenMP} threads can be seen in Figure~\ref{fig:pipipi} and Figure~\ref{fig:dmix}.

However, providing new PDFs was non-trivial, and many of the more advanced Physics PDFs had large sets of custom code that was not generalized or shared with the rest of \prog{GooFit}, such as complex return values or signal generation. \prog{GooFit} was hard to build and had specific system requirements, and the development was fragmented across several \prog{Git} repositories on \serv{GitHub}. The \prog{GooFit} 2.0 project was undertaken to make \prog{GooFit} easier to build and develop with, and to combine the development effort. Future work is addressing other aspects of the design to make PDFs easier to author and simpler to maintain.

\begin{table}
\caption{\label{tab:perf}Comparison of performance of example code from the \prog{GooFit} 2.0 examples on different systems. $\pi^{+}\pi^{-}\pi^{0}$ on the left, Zach fit on the right.}
\vspace{.1in}
\lineup
	\begin{minipage}[b]{3in}
\begin{center}
	\begin{tabular}{lll}
		\toprule
		Type & Device & Time \\
		\midrule
				2 Cores & Core 2 Duo &  \unit[1159.\0]{s} \\
GPU & GeForce GTX 1050 Ti &  \unit[\0\086.4]{s} \\
GPU & Tesla K40 & \unit[\0\064.0]{s} \\
MPI & Tesla K40 $\times 2$ & \unit[\0\039.3]{s} \\
GPU & Tesla P100 &  \unit[\0\020.3]{s} \\
		\bottomrule
	\end{tabular}
\end{center}
	\end{minipage}
\hspace{.2in}
\begin{minipage}[b]{3in}
	\begin{center}
	\begin{tabular}{lll}
	\toprule
	Type & Device & Time \\
	\midrule
		2 Cores & Core 2 Duo &  \unit[738.\0]{s} \\
		GPU & GeForce GTX 1050 Ti & \unit[\060.3]{s} \\
		GPU & Tesla K40 & \unit[\096.6]{s} \\
		MPI & Tesla K40 $\times 2$ & \unit[\054.3]{s} \\
		GPU & Tesla P100 & \unit[\023.5]{s} \\
		\bottomrule
	\end{tabular}
\end{center}
\end{minipage}
\vspace{-.1in}
\end{table}

\section{New build system}


The build system has been completely redesigned using \prog{CMake}, the most popular cross-platform build system. This was done in a three step procedure, allowing an evolutionary change to the traditional build system to support new folder structures and code reorganization while the new system was being built (described in more detail in \cite{Schreiner:2017:mg}).

The new \prog{CMake}-powered build added a plethora of new features to \prog{GooFit}. Support for IDEs, such as \prog{XCode} and \prog{QtCreator} is now available; IDE-centric features such as header listings and folder organization are included. More backends are now supported, such as the single threaded ``CPP'' backend, and a limited form of support for \comp{Intel}'s Threaded Building Blocks (\prog{TBB}). Support for \comp{Intel} and LLVM compilers was also added, including the
\comp{Apple} LLVM compiler on \oper{macOS} for the first time. Data files are automatically downloaded from \serv{GitHub} by the build system for the examples. All of the required libraries are now downloaded through \prog{Git} submodules if not discovered, including several new header-only libraries. The helper files that make these features possible are available in a separate \prog{Git} repository, for reuse by other non-\prog{GooFit} packages. A new packaging system was added as well, allowing a user to easily develop code without forking the \prog{GooFit} project.


An initial set of unit tests were added, primarily focusing on parts of \prog{GooFit} that were refactored during the 2.0 development process, such as the \comm{FitManager} and \comm{Variable}s. The examples now have a script that runs all of them sequentially, and checks the validity of the results. Continuous integration for the \lang{OpenMP} backend builds, tests, and runs the examples on the \serv{Travis} CI service. This system also builds the \prog{Doxygen} comments in the source into
documentation for every version.  Code coverage was added with \prog{GCov} and the \serv{CodeCov} service to estimate the percentage of the code base covered by the test cases, and to report changes on new pull requests.

The \serv{Travis} CI system presented several difficulties. The \prog{GooFit} library required the \prog{ROOT} library to install \cite{lib:ROOT}, but building \prog{ROOT} takes more than the alloted time on a \serv{Travis} worker node utilizing both available threads. This was overcome by using a prebuilt binary and caching it on \serv{Travis}. \prog{CMake} is trivial to download and run for any system, so a recent copy is obtained on \serv{Travis} and used as well. A standalone copy of
\prog{Minuit} 2 was made available for \prog{GooFit} 2.0, so there is longer a requirement that \prog{ROOT} be available, although since it is still used in most of the examples and a few optional tests, \prog{ROOT} remains part of the standard \serv{Travis} build.

\prog{GooFit} \prog{Docker} images for both \lang{OpenMP} and \lang{CUDA} (using \prog{NVIDIA-Docker}) were added to provide a simple way for new users to start using \prog{GooFit}. An exact set of commands to prepare a basic \prog{GooFit} environment, developed and tested using pristine operating system \prog{Docker} containers, was developed for \oper{CentOS7}, \oper{OpenSUSE}, \oper{Alpine}, and Ubuntu systems.

\section{Modernization}

\prog{GooFit} was originally designed for \lang{CUDA} 4.0 and \comp{NVIDIA} \lang{CUDA} compute architecture 2.0.  Several of the \prog{GooFit} forks were already using \Cppeleven\ features available in \lang{CUDA} 7+, so the decision was made to target \lang{CUDA} 7 and higher for the \prog{GooFit} 2.0 upgrade. Code cleanup and modernization was initially performed manually, but was slow and laborious for such a large code base. Some changes,    such as renames, were done using regular expressions in Python using the \texttt{ModernizeGooFit.py} script. This was instrumental in    removing custom terminology that would be unfamiliar to new developers. Where possible, standard \lang{CUDA} or \prog{Thrust} terminology was   used for all backends, and several spellings were normalized to match \prog{ROOT}. To facilitate rewriting \prog{GooFit} using newer     language constructs, the \prog{Clang-Tidy} program was used. It processed all of the source code using the built-in \prog{CMake} 3.6+ support, and changed a specified set of features. One feature was processed at a time, using \prog{Git} to view the     changes. Some of the most useful changes were the use of override for all overriding  virtual functions, the use of range-style          \comm{for} loops, and the use of \comm{nullptr} vs.\ \comm{0} or \comm{NULL}.

 The usage of a custom \prog{GooFit} class for complex 
 numbers was replaced with the new \prog{Thrust} complex number class. This also required supplying an external addition that enhanced the
 \texttt{ldg} \lang{CUDA} intrinsic\footnote{In \lang{CUDA}, \comm{\_\_ldg()} reads from global memory using the texture-path, which is a read-only memory path providing faster global memory access. }
 to support non-intrinsic types to retain recent performance gains on midrange \comp{NVIDIA} hardware \cite{lib:generics}.
Logging is now under a unified interface using the \prog{fmt}    library to provide simple \lang{Python}-like formatted messages \cite{lib:fmt}. Compile time settings in \prog{CMake} allow debug and trace    logging to be added to \prog{GooFit} with no runtime cost if disabled. Unified errors are provided with a custom exception subclass     that also utilizes the \prog{fmt} library. Timing statistics have been added to the standard output. A modified version of the small \prog{FeatureDetector} library  checks for missed compiler optimization and prints warnings as needed \cite{lib:FeatureDetector}.

One of the features commonly needed for the examples and for analyses was the addition of a Command Line Interface parser, \prog{CLI11} \cite{lib:CLI11}. 
This library was designed to provide a completely general utility for creating command line interfaces for complex and performance   
 dependent applications, with the ability to provide customized behavior for specific toolkits. \prog{GooFit} provides a customized subclass for the main application, adding standard \prog{GooFit} specific options, backend           information, color printing through the \prog{Rang} library \cite{lib:rang}, checks for missing compiler optimization, and a few other features. The     \prog{GooFit} version sets defaults that are designed to provide a natural one-to-one mapping of a command line
 interface and the        standard models described by fits, while \prog{CLI11} defaults remain true to standard \oper{Unix} programs.

\section{New software features}

One of the key features of \prog{GooFit} is the caching of partial results inside PDFs to be reused in the computations, taking advantage of the fact that many parameters remain the same between calls. The \comm{Variable} change detection system was improved to support multiple datasets. The caching system for specific PDFs was modified to support a ``Structure of Arrays'' format which gives better performance.

Another recent addition is preliminary \prog{MPI} support. This is included as an option in the \prog{CMake} build, and the standardized CLI11 application class allows setup and teardown to conditionally be included. The provided \prog{MPI} support divides the dataset for the application by the number of processes involved in the calculation. The compilation and execution of \prog{MPI} is supported for both \lang{OpenMP} and \lang{CUDA}. If multiple GPU processes run on the same node, they will each select a different GPU, if available.

The default fitter in \prog{GooFit} has been changed to the newly supported \prog{Minuit} 2 fitter, with the \prog{Minuit} 1 fitter still provided if \prog{ROOT} is present. The \prog{Minuit} 2 fitter was factored out of \prog{ROOT}, and a new \prog{CMake} build was added. If \prog{ROOT} is not found, \prog{GooFit} will default to the standalone \prog{Minuit} 2. As part of this change, the \prog{Minuit} 1 fitter was completely redesigned to provide a more consistent interface. The new version avoids global variables and has automated variable synchronization. The \prog{Minuit} 2 fitter provides direct access to the FCN and \prog{Minuit} 2 variables, improved logging output, and direct access to all the \prog{Minuit} 2 controls and output. 

The other non-PDF core classes in \prog{GooFit}, \comm{DataSet}s and \comm{Variable}s, were also rewritten. Input to a dataset is much simpler and faster internally, improving code transparency and maintainability. The inheritance design was improved; it is now possible to write generic code with \comm{DataSet} and select the binned or unbinned variations at runtime, simplifying several examples. \comm{Variable}s now have a much more tightly controlled API, allowing \prog{GooFit} to catch errors more reliably at compile time. Further runtime checks were added to warn unsuspecting users who load data out of range.\footnote{The author was one such unsuspecting user.} Several operator overloads were added to make manipulation of the \comm{Variable} value as convenient as direct access.

\section{New physics features}

Several physics analyses have used new physics features from the various \prog{GooFit} forks that were merged into \prog{GooFit} 2.0.

Support for three-body time-dependent amplitude analyses was added. This was used to measure mixing in the decay $D^0\rightarrow\pi^{+}\pi^{-}\pi^0$ \cite{TheBABAR:2016gom}. The code for this analysis has been provided in both \prog{GooFit} as an example, and as the original \prog{RooFit} based package. The data for this example were made public for the first time by the \babar\ collaboration just before the release of \prog{GooFit} 2.0. Another use of this feature can be seen in the LHCb mixing and CP violation search in $D^{0} \rightarrow K_S^0 \pi^{-} \pi^{+}$ \cite{Reichert:2015auv}.

Four-body time-integrated and time-dependent amplitude analyses support was added, as well. This was developed for and used in a mixing parameter search in $D^{0} \rightarrow K^{+} \pi^{-} \pi^{+} \pi^{-}$ \cite{Hasse:2017dgq}.

Toy Monte Carlo generation for multi-body systems was added using the \prog{MCBooster} library \cite{lib:MCBooster}. This is used in the three- and four-body amplitude analysis PDFs for integration, signal generation, and coordinate transformations.  Physics analyses that use this include the mixing parameter search previously mentioned as well as a model independent partial wave analysis (MIPWA) of $D^{+} \rightarrow h^{-} {h'}^{+} {h'}^{+}$ \cite{Sun:2017wtf}.

\section{GooFit future developments}

One of the most requested features, \lang{Python} bindings, has been developed as a proof-of-principle for an example, a simple exponential.  Only the minimal set of tools needed for that one example is provided in \prog{GooFit} 2.0, but it is relatively straight forward to extend to the rest of \prog{GooFit}. The bindings were constructed with \prog{pybind11} \cite{lib:pybind11}, and work with \lang{OpenMP} or GPU backends. The \lang{Python} interface is disabled by default, but is included in the test builds on \serv{Travis}.

In development for \prog{GooFit} 2.1 are drastically expanded \lang{Python} bindings, with support for most of \prog{GooFit}'s features and all of the PDFs. The current development version is available from the PyPI system through \prog{pip} (Package Installer for \lang{Python}).

\prog{GooFit} development continues at a rapid pace. A new indexing system is being designed and implemented across all PDFs, which will simplify PDF authoring and enable new optimizations to the backend in the future. The \prog{Hydra} package is being considered for inclusion, a new framework for data analysis with massively multi-threaded platforms, developed by the same author as \prog{MCBooster} as a replacement  \cite{lib:Hydra}.

\section{Conclusions}

\prog{GooFit} has undergone a major code structure transformation. The changes behind the 2.0 design is enabling new features to be added, is reducing the burden on analysts building and using \prog{GooFit}, and is encouraging contributions to a unified code base. The code runs faster and is better at catching a user's mistakes. It runs on more systems than ever and supports IDEs and debuggers. It provides a set of tools to assist analysts in writing \prog{GooFit} code. 

The future of \prog{GooFit} looks bright. The \lang{Python} bindings will make PDF composition even easier to access. The PDF indexing and redesign efforts will make developing in \prog{GooFit} easier, and could potentially improve \prog{GooFit}'s already exciting performance. \prog{Hydra} inclusion may further improve the performance and abilities of analysts.

\section*{References}
\flushleft
\bibliographystyle{iopart-num}
\bibliography{goofit20}

\end{document}